# Using Markov Boundary Approach for Interpretable and Generalizable Feature Selection

Anwesha Bhattacharyya, Yaqun Wang, Joel Vaughan, and Vijayan N. Nair

*Corporate Model Risk Wells Fargo Bank N.A.*

**Abstract**

The perceived advantage of machine learning (ML) models is that they are flexible and can incorporate a large number of features. However, many of these are typically correlated or dependent, and incorporating all of them can hinder model stability and generalizability. In fact, it is desirable to do some form of feature screening and incorporate only the relevant features. The best approaches should involve subject-matter knowledge and information on causal relationships. This paper deals with an approach called Markov boundary (MB) that is related to causal discovery, using directed acyclic graphs to represent potential relationships and using statistical tests to determine the connections. An MB is the minimum set of features that guarantee that other potential predictors do not affect the target given the boundary while ensuring maximal predictive accuracy. Identifying the Markov boundary is straightforward under assumptions of Gaussianity on the features and linear relationships between them. But these assumptions are not satisfied in practice. This paper outlines common problems associated with identifying the Markov boundary in structured data when relationships are non-linear and the predictors are of mixed data type. We propose a multi-group forward-backward selection strategy that addresses these challenges and demonstrate its capabilities on simulated and real datasets.

## 1. Introduction

Feature selection is a classical problem in statistics and data science, and it has traditionally been viewed as an essential pre-processing step for model development. However, until the 1990s, there were few applications in practice that considered large numbers of features or predictors (Guyon and Elisseeff 2003). Recent advances in computing and the advent of flexible machine learning (ML) algorithms have drastically changed the landscape. The use of subject-matter knowledge and careful (albeit computationally intensive) feature screening methods have been replaced by ML techniques that can incorporate thousands of predictors into the model upfront. The practice has

*The views expressed in the paper are those of the authors and do not represent the view of Wells Fargo.

changed to using post-hoc methods, such as permutation-based variable importance [ref], to identify the important variables after the model has been fit. There has also been an inordinate amount of emphasis on predictive performance, at the expense of model interpretability, stability, and generalizability. See, however, the pioneering work on intelligent feature selection, including by Pearl and others, discussed below.

We refer readers to [ (Guyon and Elisseeff 2003), (Yu and Liu 2004), (Wu, et al. 2013) (Bolón-Canedo, Sánchez-Maroño and Alonso-Betanzos 2015)] for excellent reviews and survey of the literature on feature selection, and discussion of model interpretability and generalizability. In particular, (Guyon and Elisseeff 2003) group feature selection methods broadly into filter, wrapper, and embedded methods, review their properties, and compare them. Filter methods are model agnostic and are used as a pre-processing step before fitting the model. Wrapper methods iterate over subsets of variables and assesses their relative importance in terms of predictive performance. Embedded methods rely on regularization of loss functions, and a popular example is Lasso (Tibshirani 1996) which utilizes an L1 penalty. However, using such forms of regularization in ML models does not guarantee inclusion of relevant features and exclusion of irrelevant features in the final model (Kohavi and John 1997).

This paper focuses on the filter method of feature selection which has the advantage that the selection algorithm is independent of the prediction algorithm and helps in effective screening of features irrespective of later models. This is useful for comparing and benchmarking different downstream models. As noted earlier, there has been considerable work on trying to identify causal features. The potential benefits of selecting these features include enhanced interpretability and robustness of predictive models. With this in mind, one possible approach is to use causal discovery algorithms with observational data (Glymour, Zhang and Spirtes 2019). These methods rely on the assumptions that the underlying probability distribution is faithful to some Direct Acyclic Graph (DAG), and the causal relationships are recovered by estimating this DAG structure. Some well-known algorithms for causal structure recovery include the PC algorithm and the FCI algorithm (Spirtes, et al. 2000). On the downside, these methods result in ambiguous causal directions (undirected edges in the graph or showing reverse directionality from what is expected) in cases where assumptions of causal sufficiency are not met and can only identify the structure up to a Markov-Equivalence class. Hence causal discovery algorithms with observational data may not be effective in many practical situations. Moreover, as highlighted in (Guyon, Aliferis and Elisseeff 2007), learning the entire causal structure to identify features useful in predicting a single target is not scalable. Nevertheless, we can still utilize the DAG assumptions to learn the local causal structure around a target variable. To be specific, these algorithms do not distinguish between direct causes or direct effects of the target but aim to learn the 'Markov boundary' of the target.

The notion of Markov blanket in the context of causal structure learning was formalized in (Pearl 2009). A Markov blanket of a target/response $Y$ was defined to be a set of variables $C$, conditioning on which all other variables were independent of $Y$. The minimal of such a set was defined as Markov boundary. Long before this, (Koller and Sahami 1996) established the optimality of using Markov Blanket for feature selection with the ultimate goal being prediction accuracy. (Tsamardinos and Aliferis 2003) elucidated the link between local causality and feature selection in faithful distributions and identified strongly relevant features in the (Kohavi and John 1997) sense to members of the Markov Blanket. The use of Markov Blanket in feature selection has been further studied extensively [(Aliferis, et al. 2010a), (Aliferis, Statnikov, et al. 2010b), (Yu, et al. 2020) and the references therein].

Markov boundary identification exploits the conditional independence relationships in the data to uncover the local structure around the target. However, testing for conditional independence is not trivial when there are non-linear relationships in the data. Furthermore, in real datasets we often find mixed data types that includes both numeric and non-numeric variables and the conditional independence tests should ideally take this into account. We have addressed these issues in this paper by developing a multi-group approach for Markov boundary identification by modifying the FBEDK algorithm proposed by (Borboudakis and Tsamardinos 2019).

The rest of the paper is organized as follows. In the method section, we introduce the Markov boundary and its connection to Directed Acyclic Graphs (DAG). We discuss the testing of conditional independence relationships that lie at the core of these DAG structures, specifically for non-linear associations. We summarize the Forward Backward Early Dropout (FBED-K) algorithm that carries out the conditional independence tests sequentially to identify the Markov boundary. This algorithm lies in the crux of our multi-group strategy which is subsequently introduced. The multi-group approach considers the non-linearity in the relationships and the numeric and non-numeric nature of the covariates and makes the selection of Markov boundary features scalable across a large feature set. In the analysis section we present our findings on simulated and real datasets.

## 2. Markov boundary and conditional independence

In this section we will define several causal concepts in the context of a Directed Acyclic Graph which will lead to the definition of Markov boundary and the assumptions required for unique identification of this set. We start by assuming that the probability distribution on a set of variables, $P(V)$ can be truly captured by a Direct Acyclic Graph (DAG) $G = (V, E)$ where $V$ is the set of nodes, which represent variables, and $E$ the set of directed edges in the graph, which represent relationships between variables. A path from $V_i \rightarrow V_j$ is a chain of direct edges stemming from $V_i$ and connecting to $V_j$ oriented in the same direction.

We introduce some of the basic concepts in a DAG. A DAG is a graph where all the edges are directed and there does not exist any cycle in the graph, i.e., there does not exist any path from $V_i$ that ends in $V_i$ for any $V_i \in V$. An edge $V_i \rightarrow V_j$ implies $V_i$ is the **parent** and $V_j$ is the **child**. If there exists a directed path from $V_i$ to $V_j$ then $V_i$ is the ancestor of $V_j$ and $V_j$ is the descendent of $V_i$. Finally, $V_i and\ V_j$ are called **spouses (colliders)** if they share a child but are not directly connected by an edge**.**

1. **Markov condition** (Pearl 2009)

*A DAG G = (V, E) and the probability distribution $\boldsymbol{P}(\boldsymbol{V})$ satisfy the Markov condition if and only if (iff) for every node $\boldsymbol{V_i} \in \boldsymbol{V},\ \boldsymbol{V_i}$ is independent of all <u>other nodes that are not its descendant</u>, i.e., ($\boldsymbol{V} \setminus \boldsymbol{Parents}(\boldsymbol{V_i}) \cup \boldsymbol{Descendents}(\boldsymbol{V_i})$) conditioning on its parents $\boldsymbol{Parents}(\boldsymbol{V_i})$.*

The triplet $(\boldsymbol{V}, \boldsymbol{G}, \boldsymbol{P})$ is called a **Bayesian network (BN)** if they satisfy the Markov condition.

The Markov condition determines a specific set of independence relations that must be true for the underlying DAG and the probability distribution. However, the probability distribution may entail additional relations (examples include lack of direct edge implies marginal independence, presence of edge implies dependence, and colliders are dependent conditioning on child). Hence, for a probability distribution to capture all relation specified by a DAG and vice-versa, we need a stronger condition than the Markov property.

2. **Faithfulness** (Pearl 2009)

*A probability distribution P and a DAG G are said to be **faithful** to each other iff all conditional independence relationships true in P are entailed by the **Markov** condition in G and vice-versa.*

If a probability distribution and a DAG are faithful to each other, then all dependent and independent relationships implied by G are true for P and vice-versa. Thus, we can use these conditional independence relations to find the set of variables which form the Markov boundary of our target. However, we need one more condition to ensure this: we have information on all variables that can explain the inter-relationships between the variables of interest. Essentially this eliminates the possibility of unmeasured confounding variables that can influence the identification of the Markov boundary. This is also known as causal sufficiency.

3. **Causal sufficiency**, (Spirtes, et al. 2000, J. Pearl 2014)].

*This condition assumes that any common cause of two or more variables in* $\boldsymbol{V}$ *is also in* $\boldsymbol{V}$ *or has constant value for all units in the population.* Here the term 'common cause' is in the sense defined in (Spirtes, et al. 2000)].

Armed with these, we can define the Markov blanket and finally the Markov boundary.

4. **Markov blanket (MBl)** (J. Pearl 2014)

*A Markov Blanket of target variable* $\boldsymbol{Y}$ *denoted as MBl(Y) is the set of variables conditioning on which all other variables are independent of* $\boldsymbol{Y}$*, that is, for every* $\boldsymbol{V_i} \in \boldsymbol{V} \setminus (\boldsymbol{MBl}(\boldsymbol{Y}) \cup \boldsymbol{Y})$, $\boldsymbol{Y} \perp \boldsymbol{V_i} | \boldsymbol{MBl}(\boldsymbol{Y})$.

It is evident that a target variable may have multiple Markov blankets including the entire set V excluding the target Y. However, the interest is in the minimal Markov blanket which is also defined as Markov boundary (MB).

5. **Markov boundary (MB)** (J. Pearl 2014)

*If no proper subset of* $\boldsymbol{Mb}(\boldsymbol{Y})$ *satisfies the property of a Markov blanket of* $\boldsymbol{Y}$*, then* $\boldsymbol{Mb}(\boldsymbol{Y})$ *is the Markov boundary of* $\boldsymbol{Y}$ *denoted as* $\boldsymbol{MB}(\boldsymbol{Y})$.

6. **Theorem 1** (J. Pearl 2014)

*Under faithfulness assumptions, the Markov boundary of a node in a causally sufficient BN* $(\boldsymbol{V}, \boldsymbol{G}, \boldsymbol{P})$ *is unique and consists of the parents (direct causes), children (direct effects) and spouses (other parents of the node's children).*

Based on Theorem 1, the MB of Lung Cancer in Figure 1 includes Smoking, Genetics (parents), Allergy (spouse), Coughing and Fatigue (children) and all other variables are independent of Lung Cancer given this set.

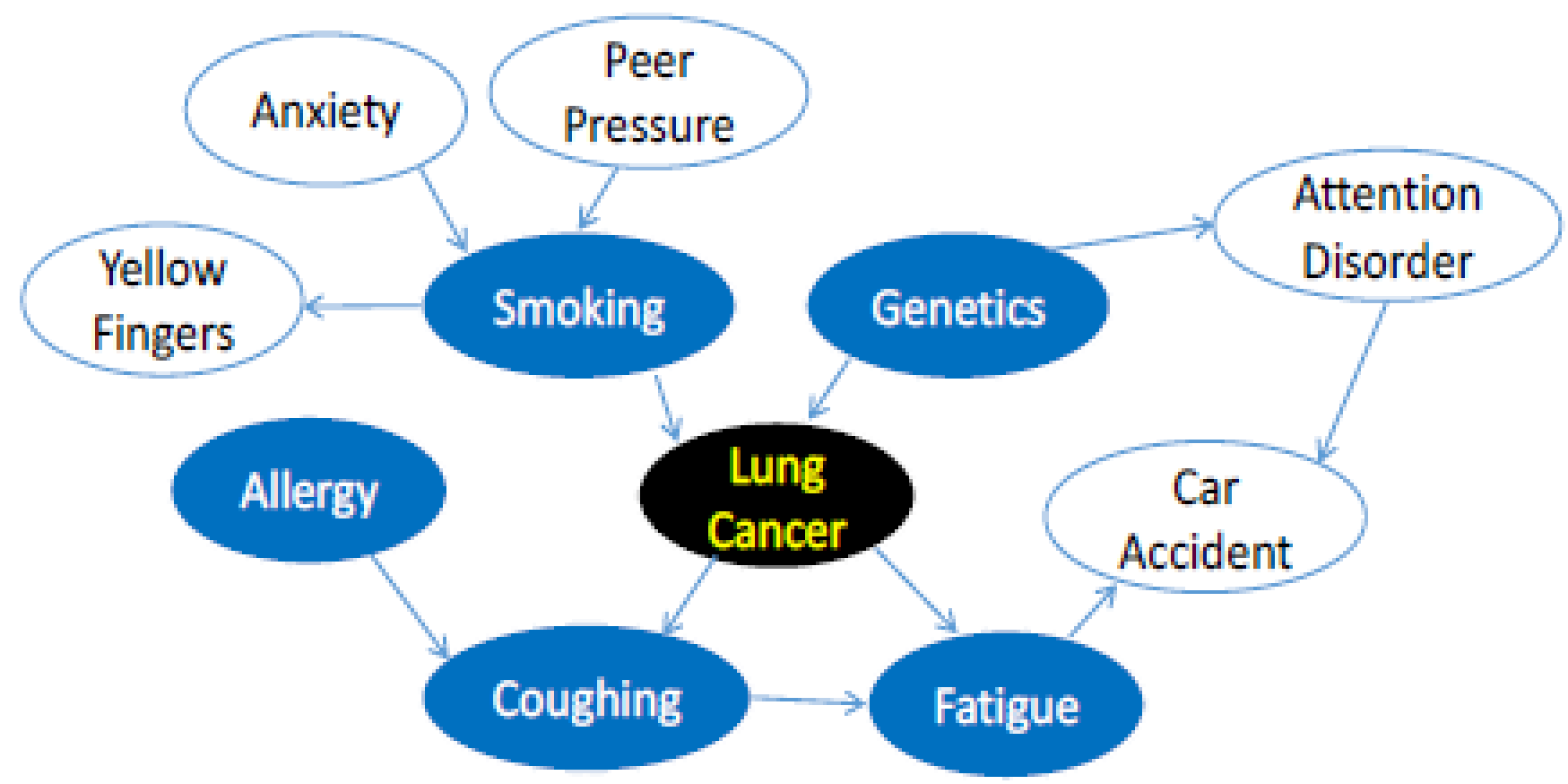

**Figure 1 Example of a Markov boundary in a DAG representing lung cancer. (Guyon, Aliferis and Elisseeff 2007) (Yu, et al. 2020)**

There has been extensive work in literature surrounding the identification of the Markov Boundary as a feature selection method. The work in (Tsamardinos and Aliferis 2003) (Yu, Liu and Li 2018) states that under faithfulness assumption, the *strongly relevant* features with respect to a variable Y (Tsamardinos and Aliferis 2003) belong to the MB(Y) and the MB is the *minimal* feature subset with *maximum predictivity* for classification.

Methods that try to identify the Markov boundary of a target variable using conditional independence relationships are known as constraint-based methods. They primarily rely on two forms of conditional independence to identify the Markov boundary. The conditions help in identification of the parents, children, and spouses that constitute the Markov Boundary of the node of interest as stated in Theorem 1.

**C.1** (Spirtes, et al. 2000) *In a BN, if* $\boldsymbol{V_i}$ *is parent or child of* $\boldsymbol{V_j}$, *then* $\forall \boldsymbol{S} \subseteq \boldsymbol{V} \setminus \{\boldsymbol{V_i}, \boldsymbol{V_j}\}, \boldsymbol{V_i} \not\perp \boldsymbol{V_j} | \boldsymbol{S}$. *S can be a null set.*

The first condition establishes the relation between a parent node and a child node. It states that as long as there is a direct edge between nodes, they will not be independent irrespective of the condition set.

**C.2** (Spirtes, et al. 2000), *In a BN if* $\boldsymbol{V_i}$ **is adjacent to** $\boldsymbol{V_j}$ **and** $\boldsymbol{V_k}$ **is adjacent to** $\boldsymbol{V_j}$ *and there does not exist a direct edge between* $\boldsymbol{V_i}$ *and* $\boldsymbol{V_k}$ *(e.g,* $\boldsymbol{V_i} \rightarrow \boldsymbol{V_j} \leftarrow \boldsymbol{V_k}$*), then* $\exists\, \boldsymbol{S} \subseteq \boldsymbol{V} \setminus \{\boldsymbol{V_i}, \boldsymbol{V_j}, \boldsymbol{V_k}\}$ *such that* $\boldsymbol{V_i} \perp \boldsymbol{V_k} | \boldsymbol{S}$ **and** $\boldsymbol{V_i} \not\perp \boldsymbol{V_k} | \{\boldsymbol{S}, \boldsymbol{V_j}\}$. *Here S is potentially an empty set.*

The second condition establishes the relationship between spouse nodes. Two nodes are said to be spouses if they do not share an edge but have a common descendant. In such a situation, the spouses are independent (marginally or conditioning on some set S not containing the descendent) but will be dependent conditioning on their common descendent(s).

To identify the MB correctly through a constraint-based method, we need to have reliable conditional independence tests for testing C1 and C2. These tests will help establish the parents, children, and spouses of the target that constitute the MB of the target under faithfulness assumptions.

## 3. Conditional Independence Tests

Since all constraint-based methods to identify Markov boundary use some form of conditional independence tests, we need a reliable test that can handle non-linear dependence. Fisher's z test based on correlation is a popular test in these methods, but these tests only work under Gaussian distribution assumption on the features. Alternatively, one could use the non-parametric Chi-Square tests of independence (Agresti 2012) and its conditional counterpart as described in (Spirtes, et al. 2000). The problem with such tests, and other methods based on binning continuous data [ (Huang 2010), (Margaritis 2005)], is that they result in data insufficiency. As the conditioning set grows (say $p$ denotes the number of variables conditioned on and $k$ is the number of categories for each variable), we need sufficient samples in $k^p$ tables, or the power of the test is sacrificed. There are other tests that rely on kernel computations and the Hilbert Schmidt criterion (see (Zhang, et al. 2011) and references there-in). Additionally, there have been some developments using distance-based correlation methods (Sz´ekely, Rizzo and Bakirov 2007) and others based on ranking in local neighborhoods (Azadkia and Chatterjee 2021). The kernel-based tests are not scalable for large datasets, while the distance correlation-based tests are only scalable for marginal dependence testing. In the methodology developed in this paper, we use the Randomized Conditional Independence test (Strobl, Zhang and Visweswaran 2018). This test is based on approximating the kernel-based conditional independence tests in (Zhang, et al. 2011) using Fourier feature approximations to speed up the test. The resulting test is scalable and Type 1 error is controlled when the Fourier approximation is good.

***Randomized Conditional Independence Test***

The Randomized Conditional Independence Test (RCIT) is motivated by the Kernel Conditional Independence Test (KCIT). The tests are scalable, and the paper offers improvement on the approximation of the null distribution (Strobl, Zhang and Visweswaran 2018) (Zhang, et al. 2011). The theoretical framework for both methods follow the characterization of conditional independence established by Fukumizu et al. in the reproducing kernel Hilbert spaces (RKHS) (Fukumizu, Bach, & Jordan, 2003).

Suppose we have $X, Y, Z$ that are continuous variables (sets of continuous variables) with domains $X, Y, Z$. We define a kernel $k_X$ on $X$ and denote the corresponding RKHS by $H_X$. Define $k_Y$, $H_Y$, $k_Z$, $H_Z$ similarly. Then for $(X, Y)$, the cross-covariance operator $\Sigma_{XY}$ from $H_Y$ to $H_X$ is defined by

$$\langle f, \Sigma_{XY} g \rangle = E_{XY}[f(X)g(Y)] - E_X[f(X)]E_Y[g(Y)], \text{ for all } f \in H_X,\ g \in H_Y.$$

Intuitively, $\Sigma_{XY}$ can be interpreted as the covariance between $\{f(X), \forall f(X) \in H_X\}$ and $\{g(Y), \forall g(Y) \in H_Y\}$.

In addition, the conditional cross-covariance operator of $(X, Y)$ given $Z$ is defined by

$$\Sigma_{XY|Z} = \Sigma_{XY} - \Sigma_{XZ}\Sigma_{ZZ}^{-1}\Sigma_{ZY}.$$

This can be interpreted as the partial covariance between $\{f(X), \forall f(X) \in H_X\}$ and $\{g(Y), \forall g(Y) \in H_Y\}$ given $\{h(Z), \forall h(Z) \in H_Z\}$. Then we can have the following characterization of conditional independence (Fukumizu, Bach, & Jordan, 2003) (Fukumizu K. G., 2007).

$$\Sigma_{\ddot{X}Y|Z} = 0 \iff X \perp Y|Z$$

with appropriate kernels and RKHSs for $X, Y, Z$ and $\ddot{X} \triangleq (X, Z)$. This can be considered as a generalization of the case when $X, Y, Z$ are Gaussian where $X \perp Y|Z$ if and only if the partial covariance of $X, Y$ given $Z$, $E[(X - E(X|Z))(Y - E(Y|Z))]$ , is zero. In other word, conditional independence is converted to un-correlatedness of residuals in the kernel spaces.

**Hypotheses**

Based on this characterization, the hypotheses for testing $X \perp Y|Z$ is set up as following in the KCIT:

$$H_0: \|\Sigma_{\ddot{X}Y|Z}\|_{HS}^2 = 0$$

$$H_1: \|\Sigma_{\ddot{X}Y|Z}\|_{HS}^2 > 0$$

where $\|\Sigma_{\ddot{X}Y|Z}\|_{HS}^2$ is the squared Hilbert-Schmidt norm of $\Sigma_{\ddot{X}Y|Z}$. $S_K = n\|\Sigma_{\ddot{X}Y|Z}\|_{\widehat{HS}}^2$ is then used as the test statistic, where $\|\Sigma_{\ddot{X}Y|Z}\|_{\widehat{HS}}^2$ is an empirical estimate of $\|\Sigma_{\ddot{X}Y|Z}\|_{HS}^2$. The calculation of the statistic in (Zhang, et al. 2011) using kernel ridge regression involves eigen decomposition of the kernel matrices that are of the order $n \times n$. Therefore, the computation scales at least quadratically with sample size, so it takes too long to complete a test when the sample size is large. RCIT tries to avoid these computations completely by approximating the RKHS with Random Fourier Features (RFF). Specially, RFF tries to construct an explicit transformation $\zeta_X: R^p \rightarrow R^d \textit{ such that } \zeta_X(x)^T\, \zeta_X(x') \sim k_X(x, x')$ where $k_X$ is a kernel function. The construction is justified by the following proposition by (Rahimi & Recht, 2007).

*For a continuous shift-invariant kernel* $k(x, y)$ *on* $R^p$*, we have*:

$$k(x, y) = \int_{R^p} e^{iw^T(x-y)} dF_w = E[\zeta(x)\zeta(y)]$$

*Where* $F_W$ *represents the CDF of* $P_W$ *and* $\zeta(x) = \sqrt{2}\cos\cos(W^T x + B)$ *with* $W \sim P_W$ *and* $B \sim Uniform([0, 2\pi])$.

Recall that in Euclidean space the Frobenius norm corresponds to the Hilbert-Schmidt norm. The hypotheses for testing $X \perp Y|Z$ can be set up, using the RFF approximation, as:

$$H_0: \|C_{\ddot{A}B|Z}\|_F^2 = 0$$

$$H_1: \|C_{\ddot{A}B|Z}\|_F^2 > 0$$

Where $C_{\ddot{A}B|Z} = E\left[(\ddot{A} - E(\ddot{A}|Z))(B - E(B|Z))^T\right]$ is the ordinary partial cross covariance matrix, $\ddot{A} = f'(\ddot{X}) \triangleq [f_1'(\ddot{X}), f_2'(\ddot{X}), \dots f_d'(\ddot{X})]$, with $f_j'(\ddot{X})\ \forall\, j$ is a RFF transformation of $\ddot{X}$, similarly $B$ for $Y$. For further details refer to (Strobl, Zhang and Visweswaran 2018). The functions $f_1', \dots f_d'$ can be intuitively thought of as basis functions approximating the RKHS associated with $\ddot{X}$. Here $\ddot{A} - E(\ddot{A}|Z)$ can be considered as the regression residual of $\ddot{A}$ on $Z$ and so does $B - E(B|Z)$ for $B$ on $Z$, so the conditional independence is converted to uncorrelatedness of residuals.

Note that $E(\ddot{A}|Z)$ and $E(B|Z)$ could be nonlinear functions of $Z$. Therefore, RCIT uses the RFF approximation trick for the kernel ridge regression in KCIT to get the residuals. Then $E(\ddot{A}|Z)$ and $E(B|Z)$ can be replaced by $E(\ddot{A}|C)$ and $E(B|C)$ where $C$ is defined similarly for $Z$ as $\ddot{A}$ and $B$. Then we have

$$H_0: \|C_{\ddot{A}B|C}\|_F^2 = 0$$

$$H_1: \|C_{\ddot{A}B|C}\|_F^2 > 0$$

**Test statistic**

Naturally, RCIT uses the following test statistic.

$$S = n\|\hat{C}_{\ddot{A}B|C}\|_F^2, \text{ where } \hat{C}_{\ddot{A}B|C} = \frac{1}{n-1}\sum_{i=1}^{n} \left[\left(\ddot{A}_i - \hat{E}(C)\right)\left(B_i - \hat{E}(C)\right)^T\right]$$

**Null distribution**

Under the null hypothesis of $X \perp Y|Z$, the test statistic $S$ has the following asymptotic distribution:

$$\sum_{k=1}^{L} \lambda_k z_k^2$$

where $\{z_1^2, \cdots, z_L^2\}$ are i.i.d. $\chi_1^2$ variables, $\{\lambda_k\}$ are the eigenvalues of the covariance matrix of the vectorization of $(\ddot{A} - E(\ddot{A}|Z))(B - E(B|Z))^T$, and $L$ is the number of elements in $\hat{C}_{\ddot{A}B|C}$. It has the same form as the null distribution in KCIT. A two-parameter Gamma distribution was used to approximate the distribution in KCIT, but the approximation was found to be crude. Therefore, in RCIT, the approximation is conducted using a finite mixture of Gamma distributions with $L$ components (the Lindsay-Pilla-Basak method) (Lindsay, 2000).

**Hyperparameters**

The numbers of Fourier features $m, q$ and $d$ for $\ddot{X}, Y$ and $Z$ are critical hyperparameters in RCIT as they control the quality of the RFF approximation. Much theoretical research has been devoted to determining values for these hyperparameters (Rahimi & Recht, 2007) (Sutherland & Schneider, 2015). In the RCIT paper, $m, q$ were set to 5 and $d$ was set to 25. Their simulation studies showed that these numbers worked well for a conditioning set size of 10 and a sample size of 1000. Our experiments (on a larger scale involving conditioning set sizes up to 20 and sample sizes up to 100K) show that $d = 25$ is not enough to control false positive rates when either conditioning set size or sample size is large. Even for $d = 100$, the false positive rate could be as high as 0.81 when the conditioning set size is 15 and the sample size is 50k. Further studies indicate a rule of thumb for the number of Fourier features to control false positive rates; 20 features per variable in the conditioning set. If the conditioning set size is 15, we would need $d = 300$ features for $Z$. Please note that increasing $d$ increases the computation time at a greater than linear rate. This motivates the multi-group algorithm detailed in the next section in which the variables are grouped to ensure that the conditioning set cannot become arbitrarily large. This not only helps to control the false positive rate but also saves computational cost. More detailed information of our empirical studies for the numbers of Fourier features is provided in Appendix A2.

## 3. Methodology

***Markov boundary identification algorithms***

Over the years, multiple algorithms have been proposed for identifying the Markov boundary of a target variable. We refer the readers to (Yu, Guo, et al. 2020) for a comprehensive review of the different approaches that have been developed. Most of the methods discussed in this review paper are based on two primary assumptions of **faithfulness (2)** and **causal sufficiency (3).**

Violations of the faithfulness assumptions lead to the existence of non-unique MBs while violations of the causal sufficiency assumption lead to incorrect identification of the Markov boundary. Although algorithms have been developed to addresses some of these issues (Yu, Guo, et al. 2020), these are hard problems, and the developed methods are either non-scalable or do not provide theoretical guarantees. Hence, we choose to proceed with algorithms that assume faithfulness and causal sufficiency.

Based on the review of (Yu, Guo, et al. 2020) and some initial experimentation we chose to work with the Forward Backward Early Dropout (**FBED**) algorithm proposed by (Borboudakis and Tsamardinos 2019). This is a simultaneous parent, child, and spouse learning algorithm which are faster than divide and conquer algorithms, but has high accuracy among the simultaneous learning class of algorithms. FBED (k = 1) also comes with the theoretical guarantee of successfully recovering the Markov Blanket if the data distribution can be faithfully represented by a Bayesian Network. The k stands for the number of sweeps on the feature set in the forward phase indicating how many times the candidate set is to be reinitialized with early dropouts.

**M1. FBED(k) algorithm** (Borboudakis and Tsamardinos 2019)

**Input**: data, target variable (Y), alpha, candidate feature set(F)

- Initiate current MB(CMB) = $\phi$, t = 0
- Forward phase: (Adding candidate features to CMB)
    - Iteration $t$, D = $\phi$
    - Select feature $X \in F$ if $F \neq \phi$ with highest association to target Y
    - If $X\ dep\ Y \mid CMB(Y)$ at level alpha
        - $CMB(Y) = CMB(Y) \cup X$
    - Else
        - $D = D \cup X$

- $F = F \backslash X$,
  - If $F = \phi$ $and$ $t < K$, initiate F with D
- Backward phase: (Eliminating false positives)
  - Select $X \in CMB(Y)$
  - If $X$ $indep$ $Y \mid CMB(Y) \backslash X$ at level alpha
    - $CMB(Y) = CMB(Y) \backslash X$

**Output:** variables in MB_set $CMB(Y)$

The FBED algorithm has the potential to grow the candidate set to arbitrarily large sizes. This poses a difficulty for the conditional independence test RCIT. For larger conditioning sets we would need to increase the number of Fourier features to get a good representation of the corresponding RKHS, but this increases the computational time significantly hampering the scalability of the algorithm. Thus, we propose a multi-group strategy to limit the size of the conditioning sets. This helps in ensuring low false positives without having to use a high number of Fourier features which would make the algorithm computationally expensive.

***Multi-Group Approach***

We rely on a grouping algorithm to divide the candidate sets to small groups such that dependent/correlated variables belong in the same group. In our implementation we use hierarchical clustering of the variables using a distance measure which is inversely proportional to the correlation matrix. The group size is controlled to be under a certain limit on average, however if the correlation threshold is too low this would still lead to large groups. In these cases, correlation threshold must be adjusted to ensure that the group size does not exceed the limit greatly. Alternately, we could compute any other measure of marginal dependence and use the inverse of this as our distance measure, or simply use a pre-specified set of groups based on domain knowledge. For each group we rely on FBED algorithms and RCIT tests to select a subset of variables in that group. As we iterate through groups, for given group, we carry out a non-parametric regression to filter out the effects of the currently selected variables from other groups on the response and use the residuals as our target. We iterate through the groups until the selected set stabilizes. In our implementation, we use an XGBoost algorithm to carry out the non-parametric regression.

**M2. Multi-group algorithm**
**Input**: data, target variable, group size, alpha, threshold

Grouping variables: continuous variables divided into multiple groups based on the group size and correlation/dependence threshold.
MB_set, the set of discovered MB variables, initialized to empty.

**Repeat:**

**For every group g:**

- Let temporary target = original target
- Let MB_set_other_groups be a set of MB variables selected from all other groups
- If MB_set_other_groups is not empty
    - use variables in the set to build a XGBoost model on original target and obtain residuals
    - update temporary target with the residuals
- Use temporary target and FBEDk algorithm with RCIT test to select variables from group g, at the alpha level.
- Update MB_set = MB_set_other_groups + selected variables from group g

**Until**

Maximum number of iterations is reached
Or early stopping if MB_set has no change.

**Output:** variables in MB_set

The multi-group approach is a convenient way to avoid tuning 'd', the number of Fourier features used in RCIT representing the RKHS. However, it has two caveats. Firstly, the XGBoost algorithm used for non-parametric regression cannot be under-fit as this would be unable to wash away the effect of the selected variables from other groups. The XGB model should not be over-fit as well, since fitting to the noise may result in unintentional washing away of signals from variables in current group. Hence the choice of the regression model requires careful consideration. Secondly, for binary response the choice of residuals is controversial. In our implementation we simply use $\boldsymbol{y} - \hat{\boldsymbol{y}}$. Simulation studies show that this choice works quite well for the proposed algorithm. However, there may be a superior choice of residuals that can work better which has not been explored in depth in this paper.

The algorithm can be further extended to take into consideration categorical variables. The categorical variables form a separate group, and we use Chi-square tests for testing conditional independence of the response (binned if it is continuous). However, an additional screening is considered before performing the FBED algorithm for each group. For each categorical variable in the group a marginal test of independence is carried out between the categorical variable and the selected continuous variables from other groups. Similarly for each continuous variable in a group, a marginal test of independence is carried out between the variable and the selected categorical variables. If a pair of categorical and continuous variable are dependent, then an additional test of conditional independence is implemented to screen out the variable from the current group(X) conditioning on the variable from the other groups (Z) ($(\boldsymbol{Y\ indep\ X|Z})$). Based on these tests we take into account dependence between categorical and continuous predictors. This screening does not guarantee accurate selection of the MB in very complex DAG structures; however, it ensures ruling out of false positives in a large number of scenarios and is shown to work well in simulation cases. The resulting algorithm is described in M3.

**M3. Multi-group algorithm with categorical inputs**

**Input**: data, target variable, group size, categorical variables, alpha, threshold

Grouping variables: all categorical variables in one group, continuous variables divided into multiple groups based on the group size and correlation/dependence threshold.

MB_set, the set of discovered MB variables. Initialized to empty set.

**Repeat:**

**For every group g:**

- Let temporary target = original target
- Let MB_set_other_groups be a set of MB variables selected from all other groups
- If MB_set_other_groups is not empty
    - use variables in the set to build a XGBoost model on original target and obtain residuals
    - update temporary target with the residuals
- If g is categorical
    - For each variable X in g test independence with MB_set_other_groups
    - If $X\ dep\ Z$ where $Z \in$ MB_set_other_groups, discretize $Z$ and test for $(X\ indep\ temp.target|discretized\ Z)$
    - If X is independent in conditional test remove X from g.
    - Use temporary target and FBEDk algorithm with Chi-square test, to select variables from group g, at the alpha level.
- If g is continuous
    - For each variable X in g test independence with MB_cat_group
    - If $X\ dep\ Z$ where $Z \in$ MB_cat _group, test for $(X\ indep\ temp.target|\ Z)$
    - If X is independent in conditional test remove X from g.

- Use temporary target and FBEDk algorithm with RCIT test, to select variables from group g, at the alpha level.
  - Update MB_set = MB_set_other_groups + selected variables from group g

**Until**

Maximum number of iterations is reached
Or early stopping if MB_set has no change.

**Output:** variables in MB_set

We demonstrate the capabilities of the algorithm in the simulation studies in next section.

## 4. Simulation Studies

The simulation study consists of two sections. In the first section we show the capability of the multi-group approach in correctly identifying the Markov boundary for three simulated datasets consisting of complex non-linear relationships and a mix of numeric and non-numeric variables for different degree of correlation between the predictors. In the second section we illustrate the role of the quality of the non-parametric regression to eliminate effect of variables between two groups and its impact on performance of the algorithm.

***Performance of Algorithm under different data generating mechanisms***

For simulation studies we look at three different datasets, a dataset which consists of linear and interaction effects on the target and a continuous set of predictors, a dataset with complex local effects and interaction effects on the target with a set of continuous predictors and finally a dataset which has complex relationships between targets and predictors and a variety of continuous and categorical predictors. The details of the data generating mechanisms are described in Appendix A1. We use a default XGB Regressor/Classifier (max_depth = 4/5, number of estimators = 200/300 and learning rate = 0.2, other HPs at default), group size = 5, and level of significance at a conservative level of 1e-4 considering Bonferroni correction. We run the simulations for 20 replicates and report the outcome in **Table 1**. The performance is measured by the F1 score on recovered features versus true features.

Table 1: Performance of M3 in terms of F1 score for different simulated datasets

| Data | Response | Correlation ($\rho$) = 0.5 | Correlation ($\rho$) = 0.8 |
|---|---|---|---|
| | | Average F1 | Average F1 |
| | Continuous | 1.000 | 1.000 |

| Linear +interactions | | | |
|---|---|---|---|
| | Binary | 1.000 | 0.997 |
| Non-linear | Continuous | 1.000 | 1.000 |
| | Binary | 0.990 | 0.970 |
| Complex | Continuous | 1.000 | 1.000 |
| | Binary | 0.993 | 0.991 |

The results show that we have good recovery in all cases with all F1 scores greater than 99%, even in the complex scenario. The algorithm is able to capture the boundary perfectly when the relationships are simple. There is usually an increase in error if the correlation between the predictors is higher or if the response is binary.

***XGBoost model used for non-parametric regression***

The efficacy of the method relies on non-parametric regression to wash out effects of selected variables from other groups in cases of weak correlation and interactions between groups. The XGBoost model used for this purpose cannot be too simplistic to not regress out relevant effects nor should it over-fit too much to noise and thereby reduce the variation in residuals, which is required to successfully capture the effects of relevant variables in the current group. To demonstrate the efficacy of this non-parametric regression we performed a simulation study with the complex data set and binary response. For weak correlation (rho = 0.2), the grouping is random, so it is possible to have correlated variables in separate groups, while for strong correlation we rely on the correlated grouping.  The results are reported in **Table 2**, with the F1 score averaged over 5 replications. The results show diminished signal for both the over-parametrized XGB model as well as the under-parametrized model for rho = 0.2. We observed that the diminished signal occurs for the spouse variables, indicating that the effect of the child was not regressed out completely in the under-parametrized model. For stronger correlation, we observe false positives in both the under and over-parametrized models. However, in all cases, the F1 score is above 90%. In general, if unsure of the complexity of the data, the hyper-parameters can be chosen based on tuning them on a section of the data. Specifically, the depth and learning rate should be tuned along with use of early stopping to get suitable number of estimators.

Table 2: Impact of using different XGB models in the algorithm reporting average of F1 score, and Recall for 5 replications.

| | | Rho = 0.2 | | Rho = 0.8 | |
|---|---|---|---|---|---|
| Model | HP | F1 | Recall | F1 | Recall |
| Default | Depth: 5<br>N_estimators: 300<br>Learning_rate: 0.2 | 1.000 | 1.000 | 0.995 | 0.991 |
| Under-parametrized | Depth: 3<br>N_estimators: 50<br>Learning_rate: 0.2 | 0.930 | 0.873 | 0.900 | 0.900 |
| Over-parametrized | Depth: 7<br>N_estimators: 300<br>Learning_rate: 0.3 | 0.995 | 0.991 | 0.944 | 0.927 |

***Comparison of M1 and M2***

We compared M1 and M2 on the three simulation datasets. As M1 cannot handle mix-type data, an all-continuous version of the complex data was used. The details of this version are described in Appendix A1. Simulation datasets were generated with two sample sizes of 50,000 and 100, 000. We run M1 with three choices of the number of Fourier features (100, 400, 800) and run M2 with its default hyperparameters: 100 Fourier features, group size of five and the default XGB regressor. For each configuration of dataset/sample size, the simulation was repeated five times with different random seeds for data generation then the average metrics were calculated, including F1 score, total number of independence tests and running time (in seconds). The results are reported in Table 3 and Table 4.

Table 3: Comparison result of M1 and M2 for sample size of 50,000

| Dataset | Method | Number of Fourier features | F1 Score | Number of independence tests | Time |
|---|---|---|---|---|---|
| Linear | M1 | 100 | 0.830 | 517.8 | 316.9 |
| | | 400 | 0.925 | 462.6 | 1028.0 |
| | | 800 | 1.000 | 417.2 | 1810.1 |
| | M2 | 100 | 1.000 | 287.0 | 214.0 |
| Nonlinear | M1 | 100 | 0.825 | 506.4 | 312.9 |
| | | 400 | 0.906 | 482.4 | 1056.8 |
| | | 800 | 0.987 | 467.6 | 2030.5 |
| | M2 | 100 | 1.000 | 296.2 | 172.6 |
| Complex | M1 | 100 | 0.671 | 510.4 | 322.4 |
| | | 400 | 0.883 | 447.8 | 987.9 |
| | | 800 | 0.982 | 419.2 | 1782.3 |
| | M2 | 100 | 1.000 | 322.4 | 189.6 |

Table 4: Comparison result of M1 and M2 for sample size of 100,000

| Dataset | Method | Number of Fourier features | F1 Score | Number of independence tests | Time |
|---|---|---|---|---|---|
| Linear | M1 | 100 | 0.830 | 522.2 | 662.5 |
| | | 400 | 0.884 | 471.4 | 2100.8 |
| | | 800 | 0.982 | 441.2 | 3844.7 |
| | M2 | 100 | 1.000 | 268.4 | 358.1 |
| Nonlinear | M1 | 100 | 0.840 | 545.2 | 660.3 |
| | | 400 | 0.853 | 509.8 | 2264.5 |
| | | 800 | 0.969 | 484.0 | 4199.4 |
| | M2 | 100 | 1.000 | 270.2 | 296.3 |
| Complex | M1 | 100 | 0.680 | 535.6 | 492.0 |
| | | 400 | 0.755 | 525.6 | 1697.1 |
| | | 800 | 0.943 | 447.0 | 2907.2 |
| | M2 | 100 | 1.000 | 290.8 | 276.5 |

The results show that M2 has good performance with perfect F1 scores in all settings while M1 has low F1 scores when the number of Fourier features are small (100 or 400). To have comparable performance as M2, M1 needs to have much larger numbers of Fourier features of 800. That leads to that much more independence tests need to be conducted so it takes much longer time to complete, about ten times longer than M1. This disadvantage of M1 makes it unfeasible in real applications when the sample size is large like several million. In such case, it would take too long time to wait for M1 having good performance. In addition, it is worth to mention that M3, the extension of M2, can handle mix-type data while M1 cannot. In practice, mix-type data is very common.

## 5. Data Analysis

We will demonstrate the utility of Markov boundary-based feature selection using two datasets including the UCI DC Bike Share Rental dataset (Fanaee T and Gama 2013) and the Taiwan Credit Dataset (Yeh and Lien 2009). For each dataset, we built a XGBoost model and a Feed Forward Neural Network model using i) a large set of predictors and ii) a selected subset of the predictors. The models were built using HP tuning on a randomly split validation set. Tuning was carried out separately for the full set of predictors and the selected set of features. For Bike Share we train models on data from year 2011 and test it on data from 2012. The results are given in Table 3. For parameter settings of the algorithm refer to Appendix A3.

It is observed that when tuned on the larger set of predictors, a more complex model is chosen. This model has a higher tendency to over-fit, producing a larger train-test gap. Reducing the number of covariates resulted in a simpler tuned model that had lower test MSE/Logloss for both the algorithms. We present the variables selected in each dataset in Table 4. In terms of variable selection, we observe that in the case of the Bike Share data, when variables are correlated the algorithms selects one of the two. For example, it selects ambient temperature from ambient temperature and temperature and season from season and month. In the case of the Taiwan data, we observe that for default prediction in the upcoming month, it uses the latest credit information and some historical information on repayment status to understand the pattern of payment in the account. We thus observe that the algorithm selects a subset of relevant variables that ensures that predictive power of the subsequent model is not diminished.

**Table 3: Predictors selected using MB identification algorithm and subsequent model performance.**

| data | Response | Set Of Predictors | Model | Metric | Train | Test |
|---|---|---|---|---|---|---|
| Bike Share | Log of count | Large set | XGB | MSE | 0.060 | 0.368 |
| | | | FFNN | | 0.128 | 0.38 |
| | | MB | XGB | | 0.117 | 0.367 |
| | | | FFNN | | 0.161 | 0.345 |
| Taiwan | Default indicator | Large set | XGB | Logloss | 0.402 | 0.430 |
| | | | | AUC | 0.818 | 0.781 |
| | | | FFNN | Logloss | 0.422 | 0.437 |
| | | | | AUC | 0.788 | 0.771 |
| | | MB | XGB | Logloss | 0.416 | 0.431 |
| | | | | AUC | 0.799 | 0.780 |
| | | | FFNN | Logloss | 0.422 | 0.432 |
| | | | | AUC | 0.790 | 0.777 |

**Table 4: Description of predictors and inclusion status in MB**

| Data | Predictors | Selected |
|---|---|---|
| Bike Share | Month | |
| | hour | Yes |
| | Holiday indicator | |
| | weekday | |
| | Workingday indicator | Yes |
| | Season (4 levels) | Yes |

| | Weather situation (4 levels) | Yes |
| --- | --- | --- |
| | temperature | |
| | Ambient temperature | Yes |
| | humidity | Yes |
| | Wind speed | |
| Taiwan | Limit balance: amount of given credit | Yes |
| | Sex | |
| | Education (4 levels) | |
| | Marriage | |
| | Age | Yes |
| | Repay status: how many months has been payment delayed at that time point.<br>Sep, Aug, Jul, Jun, May, April, 2005 | Sep,<br>Jul,<br>May,<br>April |
| | Amount of Bill Statement.<br>Sep, Aug, Jul, Jun, May, April, 2005 | Sep |
| | Amount paid in previous month<br>Sep, Aug, Jul, Jun, May, April, 2005 | Sep<br>Jul |

## 6. Discussion

In conclusion, we show that the algorithm demonstrates capability to choose an effective subset of features that helps to maintain model performance but reduces training testing performance gap. Thus, the model agnostic feature selection can help select a subset of features that lead to more robust Machine Learning models.

The approach developed in this paper identifies a relevant subset of features constituting the Markov boundary of the target using conditional independence tests that can address non-linear relationships. The approach can also handle mixed data types consisting of numeric and non-numeric features using appropriate tests. The algorithm is shown to have high accuracy in simulation setups and has successfully selected a relevant subset of features in real datasets that intuitively make sense. Subsequent Machine learning algorithms built on the subset of features demonstrate that performance is maintained in test set and gap in training and test data metrics are lower showing effective reduction in over-fitting. This suggests that the models built using the Markov boundary are more stable than models built on entire feature set. The authors report there are no competing interests to declare.

## 7. Appendix

### *A1. Simulated datasets*

**Linear with interactions**

The dataset has 51 covariates in total. The first 50 covariates are Gaussian generated from a 2-3 block covariance structure of the form

$$\Sigma = \ [\mathbf{1}\ \rho\ \rho\ \mathbf{1}\ \ \mathbf{0}\ \ \ \mathbf{0}\ \mathbf{0}\ \mathbf{1}\ \rho\ \rho\ \rho\ \mathbf{1}\ \rho\ \rho\ \rho\ \mathbf{1}\ \ \ldots\ \mathbf{0}\ \vdots \ddots\ ]$$

The last covariate $X_{50}$ is a child of the response $y$.

For continuous response

$$y = 0.6x_0 + 0.6x_1 - 0.51x_2 + 0.57x_3 - 0.57x_4 - 0.57x_5 + 0.57x_7 + 0.57x_0x_1 + 0.6x_2x_3$$
$$+ 0.7(0.6x_{10} + 0.6x_{11} - 0.51x_{12} + 0.57x_{13} - 0.57x_{14} - 0.57x_{15} + 0.57x_{17}$$
$$+ 0.57x_{10}x_{11} + 0.6x_{12}x_{13})$$
$$+ 0.4(0.6x_{20} + 0.6x_{21} - 0.51x_{22} + 0.57x_{23} - 0.57x_{24} - 0.57x_{25} + 0.57x_{27}$$
$$+ 0.57x_{20}x_{21} + 0.6x_{22}x_{23}) + \epsilon$$

$\epsilon \sim N(0,1)$

$$X_{50} = 0.2y + N(0,1)$$

For binary response

$$f(x) = -7.75 + \ 0.6x_0 + 0.6x_1 - 0.51x_2 + 0.57x_3 - 0.57x_4 - 0.57x_5 + 0.57x_7 + 0.57x_0x_1$$
$$+ 0.6x_2x_3$$
$$+ 0.7(0.6x_{10} + 0.6x_{11} - 0.51x_{12} + 0.57x_{13} - 0.57x_{14} - 0.57x_{15} + 0.57x_{17}$$
$$+ 0.57x_{10}x_{11} + 0.6x_{12}x_{13})$$
$$+ 0.4(0.6x_{20} + 0.6x_{21} - 0.51x_{22} + 0.57x_{23} - 0.57x_{24} - 0.57x_{25} + 0.57x_{27}$$
$$+ 0.57x_{20}x_{21} + 0.6x_{22}x_{23}) + \epsilon$$

$y \sim Ber(\frac{1}{1+e^{-f(x)}})$

$$X_{50} = 0.2y + N(0,1)$$

Note that we are replicating the same relationships in weaker strengths for every set of 10 predictors.

Due to the correlation in the covariates the underlying DAG structure consists of undirected edges between the covariates. The DAG consists of parents and child of the target response. A simplified version of the DAG is given in Figure 2. The boxes around nodes indicate correlated nodes. Note that this DAG structure does not consist of any spouse of the target.

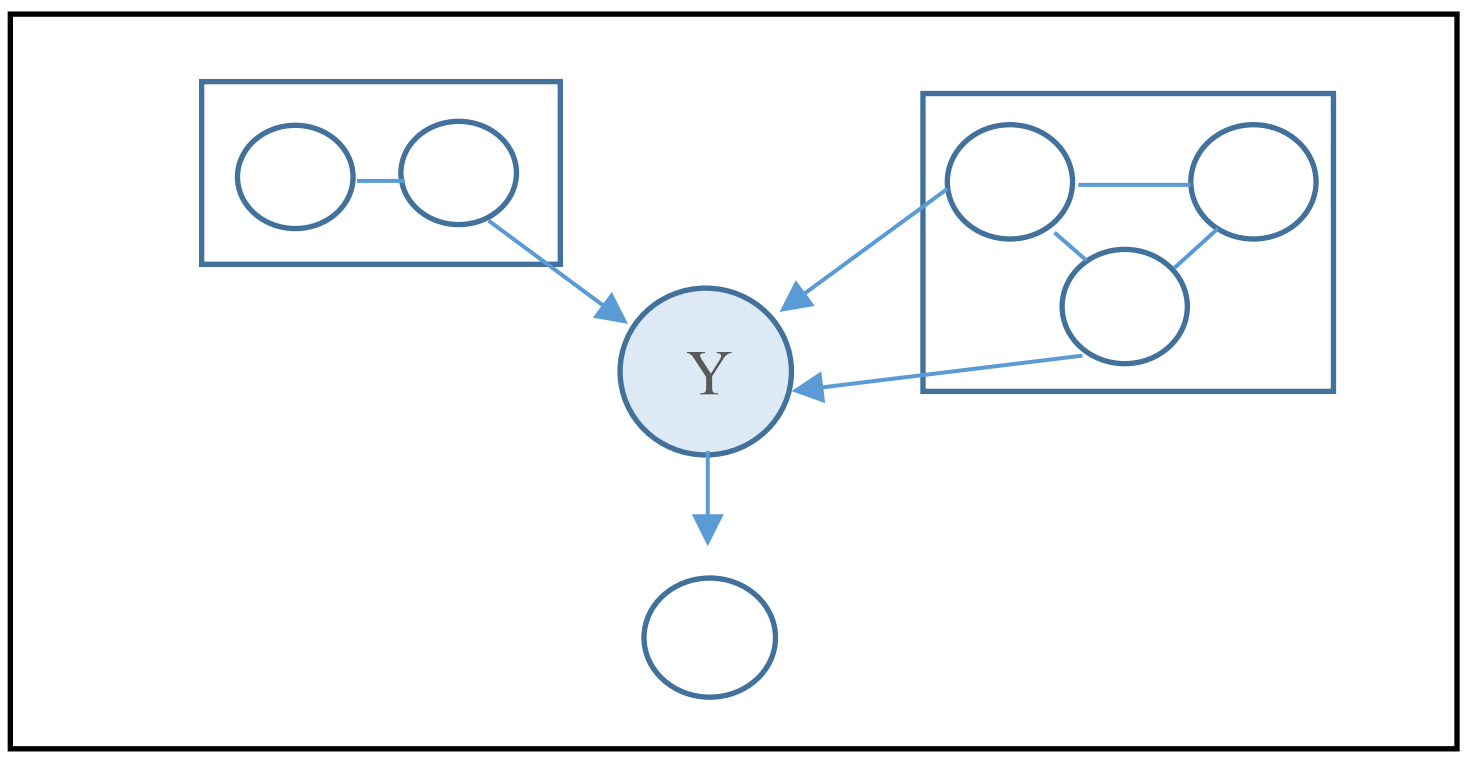

**Non-linear dataset with interactions**

The dataset has 51 covariates in total. The first 50 covariates are Gaussian generated from a 2-3 block covariance structure of the form

$$\Sigma = [\mathbf{1}\ \rho\ \rho\ \mathbf{1}\ \ \mathbf{0}\ \ \ \mathbf{0}\ \mathbf{0}\ \mathbf{1}\ \rho\ \rho\ \rho\ \mathbf{1}\ \rho\ \rho\ \rho\ \mathbf{1}\ \ \ldots\ \mathbf{0}\ \vdots \ddots ]$$

The last covariate $X_{50}$ is a child of the response $y$.

For continuous response

$$y = 0.65\, log\, (|X_0 + 0.5X_1 + 0.75X_2|) - 0.45X_0^2X_5 + |X_1\, X_2X_6| + 2X_7I(|X_7| > 2)$$

$$+\ 1.25\, X_7X_8I(X_8 < -1)$$

$$+ 0.7\left(0.65\, log\, (|X_{10} + 0.5X_{11} + 0.75X_{12}|) - 0.45X_{10}^2X_{15} + |X_{11}\, X_{12}X_{16}|\right.$$

$$\left. + 2X_{17}I(|X_{17}| > 2) + 1.25\, X_{17}X_{18}I(X_{18} < -1)\right)$$

$$+ 0.4\left(0.65\, log\, (|X_{20} + 0.5X_{21} + 0.75X_{22}|) - 0.45X_{20}^2X_{25} + |X_{21}\, X_{22}X_{26}|\right.$$

$$\left. + 2X_{27}I(|X_{27}| > 2) + 1.25\, X_{27}X_{28}I(X_{28} < -1)\right) + \epsilon$$

$$\epsilon \sim N(0,1)$$

$$X_{50} = 0.2|y| + N(0,1)$$

For binary response,

$$f(x) = -1.5 + 0.6\, log\, (|X_0 + 0.5X_1 + 0.75X_2|) - 0.45X_0^2X_5 + |X_1\, X_2X_6| + 2X_7I(|X_7| > 2)$$

$$+\ 1.25\, X_7X_8I(X_8 < -1)$$

$$+ 0.7\left(0.65\, log\, (|X_0 + 0.5X_1 + 0.75X_2|) - 0.45X_0^2X_5 + |X_1\, X_2X_6| + 2X_7I(|X_7| > 2)\right.$$

$$\left. +\ 1.25\, X_7X_8I(X_8 < -1)\right)$$

$$+ 0.4\left(0.65\, log\, (|X_0 + 0.5X_1 + 0.75X_2|) - 0.45X_0^2X_5 + |X_1\, X_2X_6| + 2X_7I(|X_7| > 2)\right.$$

$$\left. +\ 1.25\, X_7X_8I(X_8 < -1)\right) + \epsilon$$

$$y \sim Ber(\frac{1}{1+e^{-f(x)}})$$

$$X_{50} = 0.2|y| + N(0,1)$$

Note that we are replicating the same relationships in weaker strengths for every set of 10 predictors.

Due to the correlation in the covariates the underlying DAG structure consists of undirected edges between the covariates. The DAG consists of parents and child of the target response. A simplified version of the DAG is given in Figure 3. The boxes around nodes indicate correlated nodes. Note that this DAG structure does not consist of any spouse of the target.

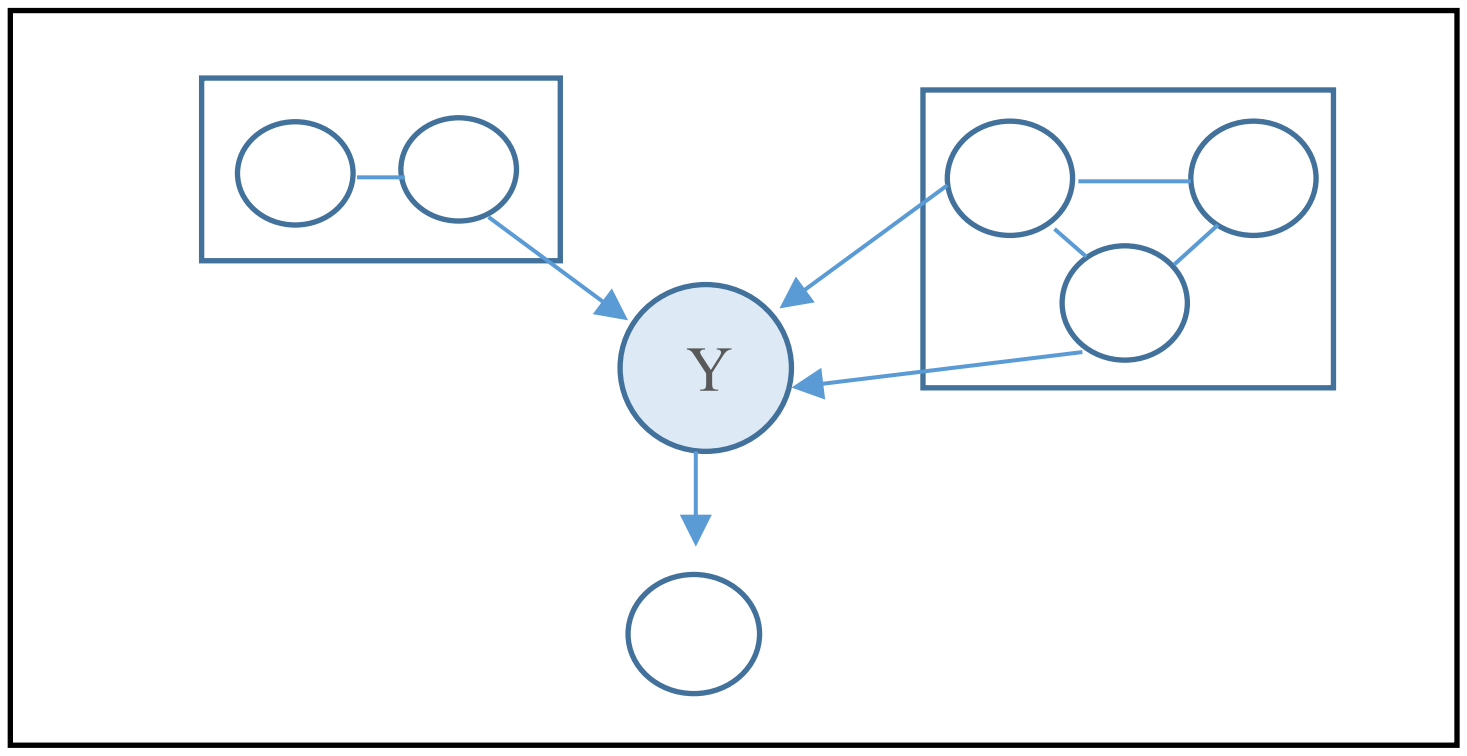

**Complex dataset**

This dataset consists of both categorical and continuous predictors. We also introduce some spousal relationships in the data. The dataset has 52 covariates in total. The first 50 covariates are initially generated from Gaussian distribution with a 2-3 block covariance structure of the form

$$\Sigma = [\mathbf{1}\ \rho\ \rho\ \mathbf{1}\ \ \mathbf{0}\ \ \ \mathbf{0}\ \mathbf{0}\ \mathbf{1}\ \rho\ \rho\ \rho\ \mathbf{1}\ \rho\ \rho\ \rho\ \mathbf{1}\ \ \ldots\ \mathbf{0}\ \vdots \ddots ]$$

We now make the following changes

- $X_{11}$ is a discrete variable formed by binning $X_{10}$. It indicates which quartile the corresponding $X_{10}$ belongs to
- $X_{12}$ is replaced by a Bernoulli indicator with probability 0.7
- $X_{13}$ is forced to be 0 whenever $X_{12}$ is 0, otherwise remains unchanged

- $X_{15}$ is replaced by an independent categorical variable that can randomly take values between 0,1,2
- $X_{30}$ and $X_{31}$ are replaced by two dependent categorical variables with two levels each
- $X_{32}$ and $X_{33}$ are replaced by two dependent categorical variables with two levels each
- $X_{36}$ is a Bernoulli variable generated from a logit function of $X_{35}$
- $X_{37}$ is a categorical variable with three levels and $X_{38}$ is a mixture normal whose mean depends on $X_{37}$
- $X_{40}$ is a categorical variable with three levels and $X_{41}$ is a mixture normal whose mean depends on $X_{40}$.

$X_{50}$ is a child of the response $y$. $X_{51}$ is a child of the response y and other covariates

For continuous response

$$
\begin{aligned}
y \ = 0.65\, log\ (|X_0 + 0.5X_1 + 0.75X_2|) \ &- 0.45X_0^2 X_5 + |X_1\ X_2X_6| + 2X_7I(|X_7| > 2) \\
&+\ 1.25\, X_7X_8I(X_8 < -1) + 0.5\, log\ (1 + |X_{11}|)\ + 0.75X_{13} + 0.5 \times 1(X_{15} \neq 1) \\
&- 0.2 \times 1(X_{15} == 2) + log\ (1 + |X_{31}|)\ + 0.75X_{32} + 0.75X_{35} + 0.5 \times 1(X_{37} \neq 1) \\
&- 0.2 \times 1(X_{37} == 1) + 0.75X_{41} + 0.75|X_{43}| + \epsilon
\end{aligned}
$$

$$\epsilon \sim N(0,1)$$

$$X_{50} = 0.2|y| + N(0,0.5)$$

$$X_{51} = 0.4y + |X_{20}| - 2\, log\ (1 + |X_{22}|)\ + e^{0.5X_{23}} + \frac{3.51}{1+|X_{25}|} + N(0,0.1)$$

For binary response,

$$
\begin{aligned}
f(x) = -5 + 0.65\, log\ (|X_0 + 0.5X_1 + 0.75X_2|) \ &- 0.45X_0^2 X_5 + |X_1\ X_2X_6| + 2X_7I(|X_7| > 2) \\
&+\ 1.25\, X_7X_8I(X_8 < -1) + +0.5\, log\ (1 + |X_{11}|)\ + 0.75X_{13} + 0.5 \times 1(X_{15} \neq 1) \\
&- 0.2 \times 1(X_{15} == 2) + log\ (1 + |X_{31}|)\ + 0.75X_{32} + 0.75X_{35} + 0.5 \times 1(X_{37} \neq 1) \\
&- 0.2 \times 1(X_{37} == 1) + 0.75X_{41} + 0.75|X_{43}|
\end{aligned}
$$

$y \sim Ber(\frac{1}{1+e^{-f(x)}})$

$X_{50} = 0.2y + N(0,0.5)$

$X_{51} = 0.4y + |X_{20}| - 2\ log\ (1 + |X_{22}|)\ + e^{0.5X_{23}} + \frac{3.51}{1+|X_{25}|} + N(0,0.1)$

Due to the correlation and other forms of dependence in the covariates the underlying DAG structure consists of undirected edges between the covariates. The DAG consists of parents, children and spouses of the target response. A simplified version of the DAG is given in Figure 3. The boxes around nodes indicate interdependent nodes.

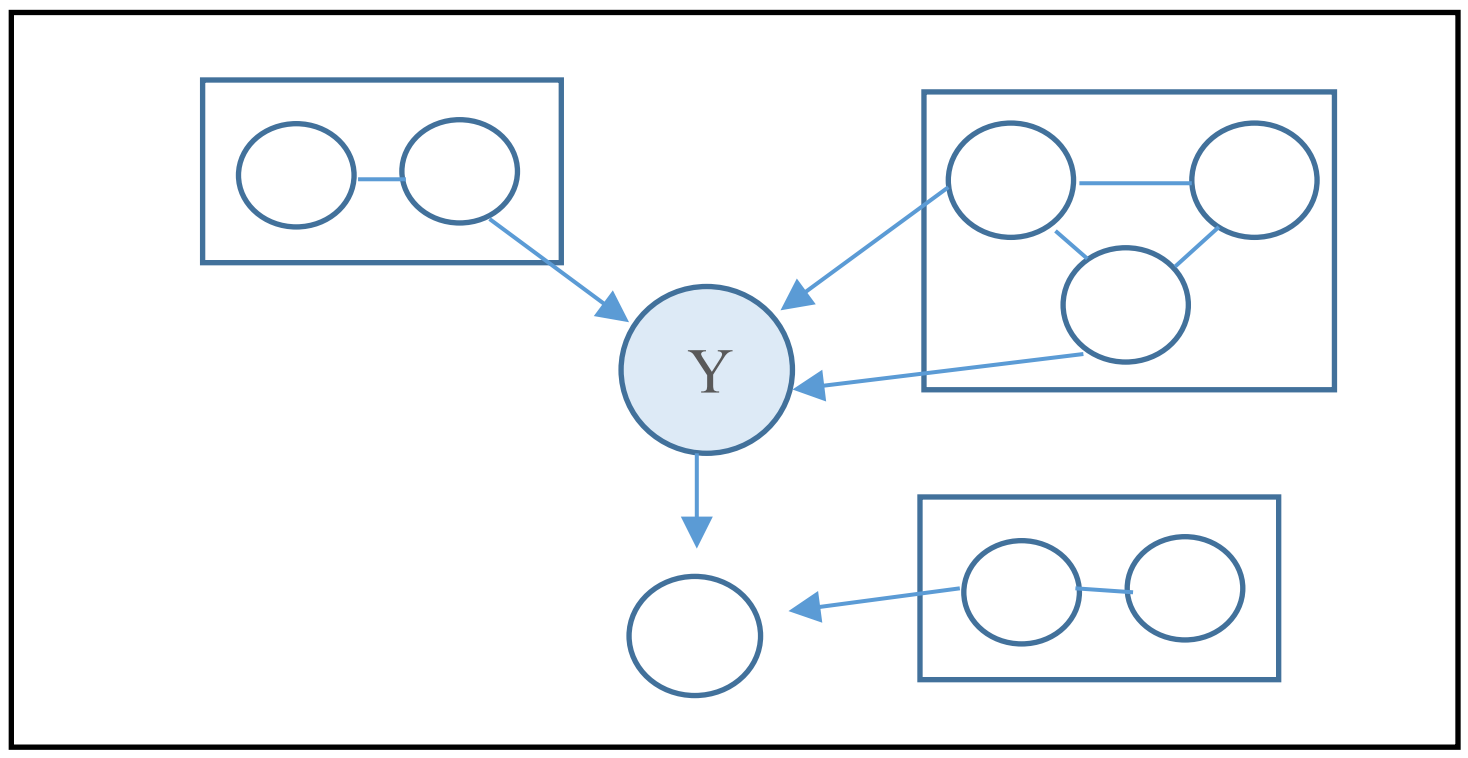

*A2. Empirical study of RCIT*

We conduct empirical experiments to investigate how many random features are needed in RCIT to ensure false positive rates are controlled. Our experiment is at a much larger scale than the scales mentioned in (Strobl, Zhang and Visweswaran 2018) with conditioning set sizes up to 20 and sample sizes up to 100k. Specifically, we use a simulated dataset similar to the Complex dataset in A1 but only including continuous variables. In the simulation, variables $X_{37}$, $X_{38}$ and $X_{39}$ are highly correlated but only $X_{37}$ is used to generate the response $y$. We test the independence between $y$ and $X_{39}$ given different conditioning sets with different sample sizes. The variable $X_{37}$ is always included in the conditioning sets along with different numbers of other variables. We expect the test result to be negative as $X_{39}$ doesn't provide additional information for the response conditional on $X_{37}$. Therefore, any positive test result is considered as false positive. For each setting, we repeat the test for 100 times with 100 randomly generated datasets to calculate the false positive rate. In Table 5, we can see that, even for $d = 100$, the false positive rate could be as high as 0.81 when the conditioning set size is 15 and the sample size is 50k.

Table 5 False positive rate and computation time for different values of $d$

| **n** | **Conditioning Size** | $d$ | **False Positive** | **Time** |
|---|---|---|---|---|
| 50000 | 15 | 100 | 0.81 | 8.99 |
| 50000 | 15 | 200 | 0.17 | 23.41 |
| 50000 | 15 | 300 | 0.01 | 50.95 |
| 50000 | 15 | 400 | 0.00 | 86.19 |
| 50000 | 15 | 500 | 0.00 | 131.81 |
| 50000 | 15 | 600 | 0.00 | 187.75 |

In a further investigation, we study the false positive rate for different numbers of Fourier features per variable in the conditioning set. We discover that 20 features per variable is sufficient for controlling the false positive rate (see Figure 5).

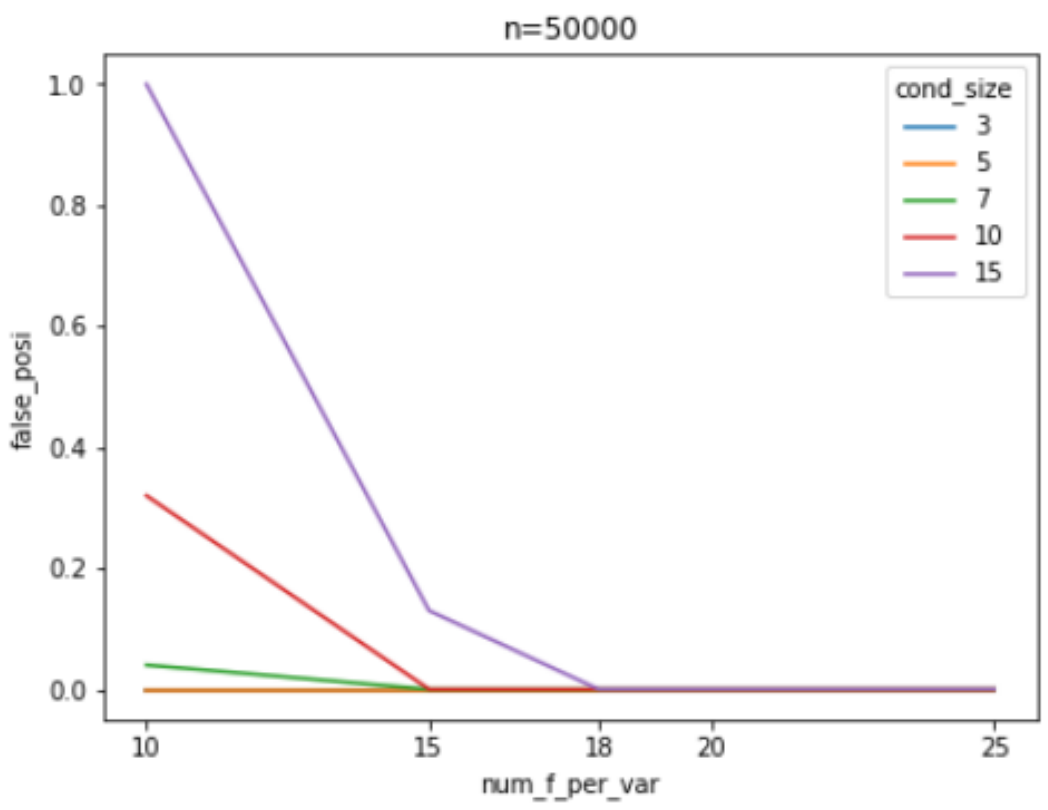

**Figure 5 False positive rate for different numbers of Fourier features per variable in the conditioning set, for different conditioning set sizes**

If the conditioning set size is 15, we would need $d = 300$ features for $Z$. However, we can see from Table 3. that the increasing of $d$ would increase the computation time at a greater than linear rate. For example, the computing time is about 9 seconds when $d = 100$ but increases to 188 seconds when $d$ is six times larger.

### *A3. Data Analysis*

We grouped variables with correlation greater than 0.2 together. Due to the large number of correlated variables in the Taiwan data, we used 400 Fourier features for RCIT, whereas for Bike Share only 100 Fourier Features were used. The algorithm was run till the feature set converged. We used 1e-6 as a conservative level of significance accounting for multiple testing in the algorithm. The XGB and FFNN algorithms were tuned in the same grid space for the full model and the sub model.